\documentclass[mathleft
]{an}
\usepackage{graphicx}
\usepackage{times}
\def\h{{$^{\rm h}$}}
\def\m{{$^{\rm m}$}}
\def\s{{$^{\rm s}$}}
\begin{document}

\Pagespan{1}{}
\Yearpublication{2011}%
\Yearsubmission{2011}%
\Month{9}%
\Volume{xxx}%
\Issue{xx}%
\title{The sunspot observations by Samuel Heinrich Schwabe}
\author{R. Arlt\thanks{Corresponding author:
  {rarlt@aip.de}}
}
\titlerunning{Sunspot observations by Schwabe}
\authorrunning{R. Arlt}
\institute{
Leibniz-Institut f\"ur Astrophysik Potsdam, An der Sternwarte 16, 
D-14482 Potsdam, Germany
}

\received{...}
\accepted{...}
\publonline{later}

\keywords{Sun}

\abstract{%
  A long time-series of sunspot observations is preserved
  from Samuel Heinrich Schwabe who made notes and drawings
  of sunspots from 1825--1867. Schwabe's observing records 
  are preserved in the manuscript archives of the Royal Astronomical
  Society, London. The drawings have now been digitized for 
  future measurements of sunspot positions and sizes. The 
  present work gives an inventory and evaluation of the images 
  obtained from the log books of Schwabe. The total number of
  full-disk drawings of the sun with spots is 8486, the number 
  of additional verbal reports on sunspots is 3699. There
  are also 31~reports about possible aurorae.
  }

\maketitle

\section{Introduction}
Understanding the origin of the solar dynamo requires an extensive
time series, since the cycle is not fully periodic, and only a large
number of cycles can reveal the statistical properties produced by
the dynamo mechanism. Solar activity is well covered by sunspot
observations since 1874 when the Greenwich data start. Smaller sets
are available from Gustav Sp\"orer (Sp\"orer 1874, 1878, 1880, 1886, 1894) 
and Richard Carrington (1863). A large series of several thousand 
observations was recorded by Samuel Heinrich Schwabe in the period
of 1825--1867. Together with the observations by Sp\"orer, a complete
series of sunspot positions from 1825 to today can be constructed.
Note however, that Sp\"orer published only a few positions for sunspot 
groups instead of all individual spots. Tilt-angles of sunspot groups 
cannot be retrieved from his tables. He did publish drawings of the
groups at some evolved stage, whence sunspot areas can potentially
be determined.
Schwabe provided full-disk drawings, however. It will thus be extremely 
interesting to measure the positions of the individual sunspots plotted in these 
drawings and determine the sunspot sizes as well.

Schwabe was born in Dessau on 1789 Oct~25 and started with his astronomical
observations rather late in 1825, at an age of almost 36. While supposed to
take over the pharmacy from his grandfather, he went to Berlin for the
corresponding studies, and even joined university to listen to courses 
on experimental chemistry, botanics, and experimental physics. Schwabe 
did no finish the studies because he had to go back to Dessau in December 
1811 and, being the eldest son, to take care of his family (Arendt 1925).
He died in Dessau on 1875 Apr~11. Schwabe obtained the Gold Medal of the Royal
Astronomical Society of 1857. Richard Carrington very likely handed the
Medal over to Schwabe already when he visited him on 1856 Oct~15, according 
to Schwabe's notes, in which the latter did not mention the Medal though.
According to a letter from Schwabe to Haase written in 1862, it was
C.L.~Harding (co-editor of ``Kleine Ephemeriden'') who initially motivated 
Schwabe to carry out sunspot observations with the aim to find a possible 
planet inside the orbit of Mercury (Arendt 1925).

Extended activities in botanics go back to Schwabe's childhood, and he
published a seminal systematics of the flora in his region in the
``Flora Anhaltina'' in 1838.

\section{The log books}
The observing books of Schwabe are stored in the library of the
Royal Astronomical Society in London and are in very good
condition. In 1864, representatives of the RAS inquired
to obtain the astronomical observations from Schwabe. He 
agreed under the condition that he can get them back whenever
he wishes to carry out further analyses (being at an age 
of 75). He added: ``After my death you may consider the
whole of the Observations as the property of the Royal
Astronomical Society'' (Huggins 1876).

All observations were recorded in note books with a binding.
There is no possibility of disordered lose sheets of papers.
The total number of books is 39 with a few books containing
the information of two years (Bennett 1978). 
The full period of time covered by solar observations spans from
1825 Oct~1 to 1867 Dec~31. The first drawing of the solar disk
was made on 1825 Nov~05. In the beginning, until January 1826,
several observations were combined in a full-disk drawing
while it is always clearly mentioned which spots refer to which
date. The last drawing was made on 1876 Dec~29,
while the last verbal report is from 1867 Dec~31. The records
stop without any mention of the reasons, but since several of
the observations in the last years were actually recorded by
people assisting Schwabe, we may assume that his health did not 
allow any further observations.

\begin{figure}
\begin{center}
\includegraphics[width=0.485\textwidth]{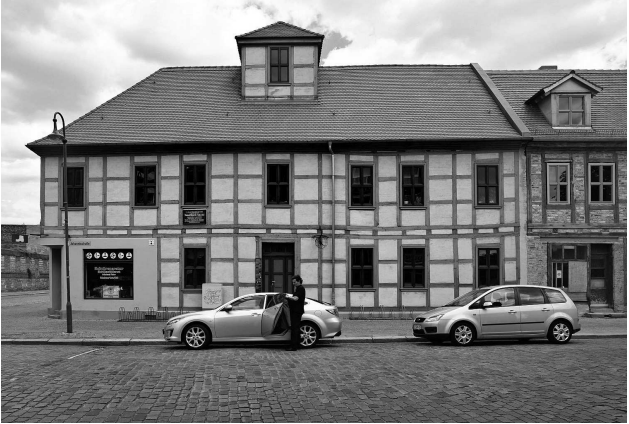}
\end{center}
\caption{The living house of Samuel Heinrich Schwabe and his family 
in Dessau, photographed on May~29, 2011. Most of the observations
were made from the turret on top of the building.}
\label{schwabehaus}
\end{figure}

The geographical coordinates of the observing location in
Johannisstra\ss{}e 18, Dessau (address still valid), are 
$\lambda = 12\degr 14'32''$\,E, $\phi = 51\degr 50'19''$\,N 
(WGS84) from where he started observing on 1830 May~17. A 
contemporary photo of the restored building is shown in 
Fig.~\ref{schwabehaus}. The observing tower on top of it
may not be in the original style, and Schwabe actually changed
the windows a few times to accommodate his telescopes and a
transit instrument, as indicated by some of his notes, e.g.
``In the evening, I had a hole broken into the wall and corrected
the orientation of the cross hairs'' (1840 Jun~20).

\begin{figure}
\begin{center}
\includegraphics[width=0.485\textwidth]{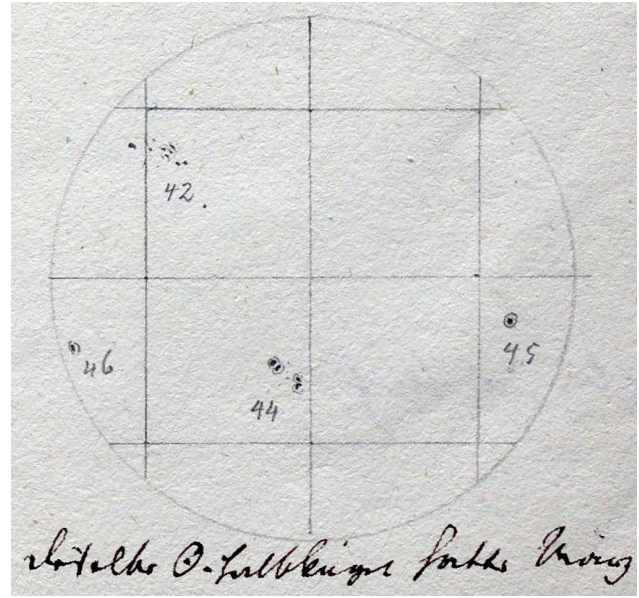}
\end{center}
\caption{Drawing of the sun of 1847 Apr~14.}
\label{grid1}
\end{figure}

All full-disk drawings were made by pencil. On some pages,
especially where the drawings are near the lower edge of the
pages, the clarity of the pencil marks are somewhat washed
out, but nowhere to the extent that information about sunspots
would be lost. Schwabe assigned numbers to the sunspot groups
which were added after the observation in ink or by pencil.
An typical example of such a drawing is shown in Fig.~\ref{grid1}
from the observation of 1847 Apr~14.





\begin{figure}
\begin{center}
\includegraphics[width=0.485\textwidth]{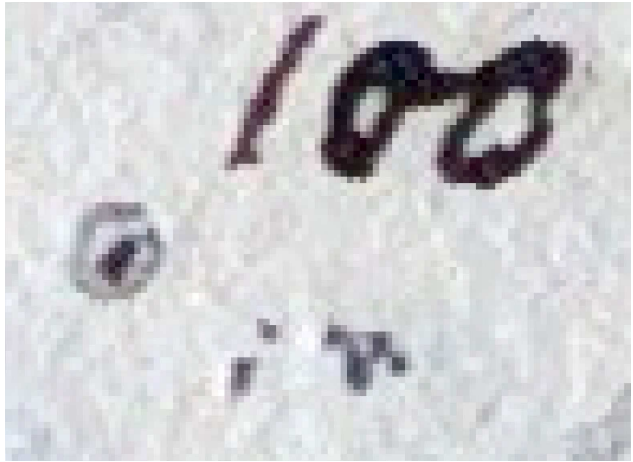}
\end{center}
\caption{Enlargement of a part of a full-disk drawing showing
the pixel accuracy of the digitized images.}
\label{enlargement}
\end{figure}

The logbooks were digitized photographically. Full pages
were photographed with a Canon EOS 5D with a resolution of 
$2912\times 4378$~pixels. The lens was a Sigma 50mm/2.8 EX DG 
Macro which provides an extremely low image distortion. The 
image scale is about 0.07--0.08 mm/pixel, depending on the 
sizes of the books. The images were taken with apertures 
of f/8--f/9 with ambient light from a near window, since 
the inclined infall of light preserves the paper structure 
and pencil engravings very well. This illumination will later 
help to distinguish the actual sunspots from paper defects and 
ink spots. All images were taken in the proprietary raw, 12-bit image
format by Canon and processed automatically in RawTherapee~2.3 
of February 2008\footnote{http://www.rawtherapee.com/}.
Figure~\ref{enlargement} shows an enlargement of an example
image of 1838 May~02 without interpolation to show the
pixel scale of the digitization. Even small spots are 
represented by several pixels to have enough accuracy for
the positional measurements.

The images of the full pages were then cut into images for
individual days as it was done for the drawings by Staudacher
in Arlt (2008). Usually, a strip containing the drawing on
one side and the description on the other side is cropped
from the full-page image. Sometimes the descriptions are so
lengthy that only an abridged version is visible in the
final image. File names were constructed with the date and time
in the format {\tt YYYY\_MM\_DD\_HHMM.JPG}. If there is no time
in Schwabe's description which can be associated with the 
drawing, we set {\tt HHMM} to {\tt 0000}.

\section{Types of information}
The majority of sunspot observations consists of a circle with
the distribution of sunspots. The circles have a diameter of about 4.5~cm.
There is a considerable amount of detailed drawings
of selected sunspot groups, but they are concentrated in the
beginning of the entire observing period. Apart from observations 
with drawings, we find a lot of dates for which just the changes 
to the previous day was described verbally. These observations
can be interesting for investigations of the life-time of 
sunspots, since Schwabe noted which sunspots had disappeared
and if a new sunspot had appeared, mostly during the afternoon
hours.

The last group of observations are about days on which no sunspots 
were seen. These were in fact the observations which led to the 
discovery of the solar cycle. Schwabe simply counted the days 
without sunspots per year and showed the variability of the 
(inverse) solar activity according to the 18~years of 1826--1843 
(Schwabe 1844). He actually wrote the letter to Astronomische 
Nachrichten on Dec~31, 1843, immediately when he completed his 
set of data capable of proposing the cycle.

\begin{figure*}
\begin{center}
\includegraphics[width=0.34\textwidth,angle=90]{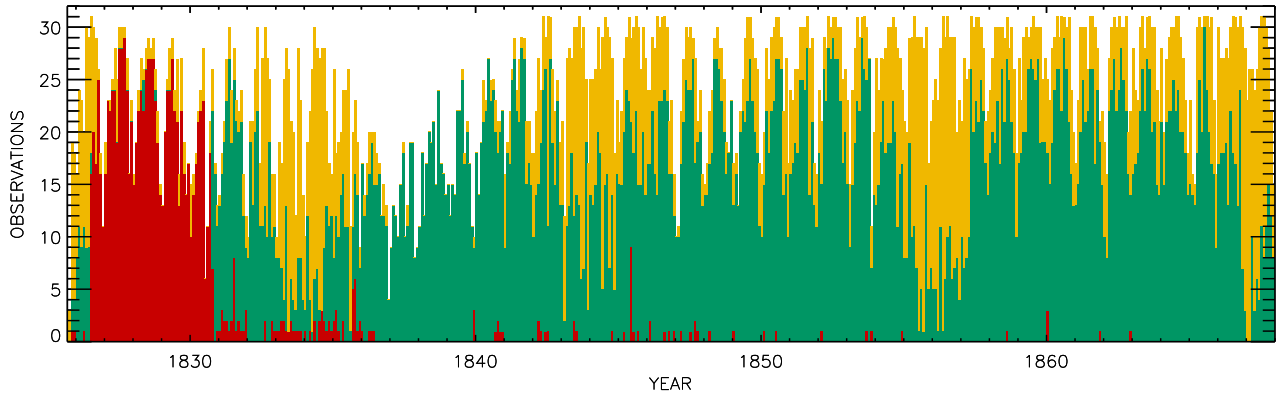}
\end{center}
\caption{Distribution of observations per month over time. The
yellow (light) bars show the total number of observations per month,
including verbal reports. The green (darker) bars depict the number
of drawings with sunspots, while the red (darkest) bars show the
number drawings which have no coordinate system.}
\label{schwabe_stat}
\end{figure*}

A total inventory of Schwabe's sunspot observations is given
in Table~\ref{inventory}. The numbers of drawings and the numbers 
of verbal reports without drawings are shown. We also list the
numbers of drawings which show no ``coordinate system'' and the
numbers of drawings which cannot be associated with a specific
time of day (see Sect.~\ref{times}). The inventory is also shown
graphically by monthly totals in Fig.~\ref{schwabe_stat}. One
can see that there is actually a large number of months in which
Schwabe managed to report on sunspots on {\it every\/} day. The
annual variation comes from the fact that the winter months have
fewer sunny days.

Schwabe very often reported on faculae for which he used the
German word construction ``Lichtgew\"olk" corresponding to ``light
clouds''. Since he had to indicate these bright structures with
a grey pencil, it is often not easy to distinguish faculae from
sunspots in the drawings. The verbal information as well as the
absence of group numbers (for faculae without spots) are helpful 
here.

The limb darkening was also noted on several occasions, specifically
in 1841, 1846, 1847, and 1851, while in 1867 Schwabe noted that the
limb was not darker than the rest of the disk. He also reported about
the appearance of the granulation a few times and describes it as
``the marbled surface of the Sun'' (1841 Apr~05).

Many other astronomical observations accompany the sunspot drawings 
and descriptions, most notably of the Moon, Jupiter, and Saturn. Schwabe
commented on meteors a few times, described a few fireballs, and gave 
descriptions of solar and lunar halos, rainbows and 31~occasions of aurorae. 
The latter are rather frequent given the relatively low latitude of 
his observing place. The descriptions resemble aurorae quite well,
however, and they seem to cluster near or shortly after solar maxima.
A list of Schwabe's aurorae sightings is given at the end of this
paper in Table~\ref{aurorae}.
 
\begin{table}
\caption{Annual numbers of sunspot observations by Heinrich Schwabe. The total
number of observations is 12185.\label{inventory}}
\begin{tabular}{lrrrr}
\hline
Year & Drawings & Drawings  & Drawings & Verbal  \\
     &          & w/o time  & w/o grid & reports \\
\hline
1825 &  24      &  0 &   1 & 14 \\
1826 & 165      &  3 & 106 &121 \\
1827 & 274      & 38 & 274 & 25 \\
1828 & 269      & 61 & 265 & 20 \\
1829 & 233      & 45 & 233 & 26 \\
1830 & 191      & 18 & 150 & 26 \\
1831 & 225      & 64 & 27  & 34 \\
1832 & 167      &  7 &  5  & 99 \\
1833 &  74      &  1 & 14  &169 \\
1834 &  84      &  2 & 15  &189 \\
1835 & 131      &  4 & 22  &108 \\
1836 & 163      &  4 &  4  & 24 \\
1837 & 153      & 41 &  0  &  3 \\
1838 & 204      & 26 &  0  &  0 \\
1839 & 199      & 30 &  3  &  4 \\
1840 & 248      & 13 &  5  & 15 \\
1841 & 241      & 14 &  0  & 40 \\
1842 & 216      &  5 &  5  & 89 \\
1843 & 131      &  4 &  3  &177 \\
1844 & 158      &  3 &  1  &161 \\
1845 & 222      &  3 & 12  &109 \\
1846 & 223      &  7 &  5  & 96 \\
1847 & 227      & 22 &  5  & 49 \\
1848 & 243      &  5 &  1  & 35 \\
1849 & 227      &  4 &  1  & 58 \\
1850 & 229      &  4 &  2  & 77 \\
1851 & 231      &  4 &  0  & 79 \\
1852 & 280      &  5 &  1  & 56 \\
1853 & 231      &  3 &  1  & 69 \\
1854 & 213      &  2 &  1  &120 \\
1855 &  97      &  0 &  0  &216 \\
1856 &  75      &  0 &  0  &245 \\
1857 & 203      &  3 &  0  &120 \\
1858 & 245      &  0 &  1  & 88 \\
1859 & 263      &  1 &  0  & 79 \\
1860 & 255      &  1 &  3  & 77 \\
1861 & 243      &  0 &  1  & 76 \\
1862 & 242      &  0 &  1  & 75 \\
1863 & 244      &  0 &  0  & 87 \\
1864 & 244      &  0 &  0  & 79 \\
1865 & 221      &  0 &  0  & 86 \\
1866 & 189      &  5 &  0  &161 \\
1867 &  88      &  0 &  0  &224 \\
\hline
Totals & 8486   &    &1168&3699\\
\end{tabular}
\end{table}

Schwabe observed several comets and asteroids after their 
discoveries, and he found Neptune on 1846 Sep~26, three days 
after its discovery. Le Verrier predicted another planet inside
the orbit of Mercury on 1860 Jan~2 for which Schwabe also looked
during his solar observations. There are many notes that nothing
was seen except on 1862 Aug~06, he writes: ``Is [group] 101 perhaps
a planet between Mercury and the Sun, but at 12h there was no notable
motion.'' He also mentioned the visibility of the zodiacal 
light several times in 1843--1855. Schwabe reported about the
capability of resolving a number of binary stars during the course of
the years. These observations allow us to judge about the quality
of his telescopes.

Schwabe also noted the temperatures for the morning, afternoon,
and evening of each day since 1827 Jan~21. Barometric measurements
were started on 1832 Jan~01. Information about the wind direction
for the same three instances as for the temperature and the pressure
starts on 1852 Oct~01.

\begin{table*}
\caption{The telescopes of Samuel Heinrich Schwabe sorted by the
order in which they appear in the logbooks.\label{telescopes}}
\begin{tabular}{llll}
\hline
Focal length & Manufacturer & First mention & Remarks\\
             & or Schwabe's designation if unknown &         &        \\
\hline
2 1/2 foot   & Ramsden              & 1825 10 01 & \\
1 1/2 foot   & Thomas Harris \& Son & 1825 10 07 & \\
2 1/2 foot   & Winkler              & 1825 11 22 & more precisely 29~inch\\  
3 1/2 foot   & Fraunhofer           & 1826 01 22 & \\
Sternsucher  &                      & 1826 10 23 & \\
18 inch      & Cometensucher        & 1827 07 05 & \\
7 inch       & Taschenperspectiv    & 1827 07 05 & possibly identical to ``Sternsucher''\\
5 inch       & Cometensucher        & 1828 09 04 & \\
16 inch      & Cometensucher        & 1828 09 04 & \\
17 inch      &                      & 1828 10 15 & 16, 17, 18 inch possibly identical\\
6 foot       & Fraunhofer           & 1829 02 13 & obtained from Lohrmann, Dresden\\
20 inch      & Bobbe                & 1840 06 20 & also called ``Mittagsfernrohr''\\
2 1/2 inch   &                      & 1853 08 16 & most likely identical with Winkler's\\
\hline
\end{tabular}
\end{table*}

\section{Schwabe's instruments}
The drawings are most likely aligned with the equatorial
system, with north pointing downward and east pointing to the 
right. This is the view through a Keplerian telescope. For
example, the description of 1840 Apr~18 says that new spots 
`entered' the disk and these were drawn on the right side of 
the circle drawn. There are numerous of such verbal information
all being consistent with east being on the left side of the
image. 

The images could still be mirrored as the result of a projection 
method, but the same description of 1840 Apr~18 also refers to 
another spot `south' of the previous one, which is actually 
plotted above the previous one. On 1838 Apr~12, Schwabe mentions
``spots only in the north'' while the drawing shows spots only
in the lower half. The image must thus be a rotated version of the 
solar disk as seen in a Keplerian telescope, and not a mirrored one. 

Already in the very first observation of 1825 Oct~1, Schwabe noted 
that ``the absorption glass appears to be too dark.'' Several other
verbal comments deal with dimming glasses for observations
through the telescope, as the following translated phrases
illustrate:
\begin{itemize}
\item 1826 02 01: ``During the observation the sun glass broke''
\item 1826 02 02: ``through a lava sun glass by Winkler''
\item 1826 03 23: ``with the brightest sun glass by F[raun\-ho\-fer]''
\item 1826 05 02: ``with the sun glasses by Fraunhofer''
\item 1826 11 29: ``I could only use the sun glass by Utzfohl'' -- manufacturer not verified
\item 1827 07 24: ``with all my sun glasses''
\item 1854 07 16: ``with the old, usual sun glass''
\item 1854 10 28: ``the violet glass from Leipzig'', ``the pale-blue glass from Munich''
\end{itemize}

We therefore assume that he was looking through Keplerian 
telescopes with filters for the entire observing period.

Schwabe used several telescopes during his 43~years of 
astronomical observations. The first telescope mentioned
is a ``2-1/2-foot Ramsden'' with an eyepiece magnifying 
70 times (1825 Oct~01) and unknown aperture. The sizes 
refer to the focal lengths of the telescopes. In parallel, 
he used a ``1-1/2-foot telescope by Harris'' of which 
aperture and magnification are unknown (1825 Oct~07). The 
company Thomas Harris \& Son, London, produced telescopes 
with diameters typically between 4~and 6~cm between 
1806 and 1846. Schwabe obtained his third telescope on
Nov~22, 1825, when he wrote ``I observed with my 2-1/2-foot
telescope by Winkler which I received from Leipzig today.''
This telescope was also mentioned by Arendt (1925).

Schwabe eventually received a 3-1/2-foot telescope from Fraunhofer 
on 1826 Jan~22, and the first drawing made with this is probably
from 1826 Jan~29. This telescope was used for the vast majority
of full-disk drawings made from 1826--1867. A second Keplerian 
telescope with 6~feet focal length and 12.2 cm aperture from 
Wilhelm Gotthelf Lohrmann in Dresden, Germany, was received on 
1829 Feb~13 and used for solar observations since 1829 Feb~26. 
The lenses of the telescope were also manufactured by Fraunhofer 
while the construction and parallactic mount was made by 
Rudolf Sigismund Blochmann (Arendt 1925; Weichold 1985). These 
and additional telescopes, which have little relevance for 
the sunspot observations, are listed in Table~\ref{telescopes}.

\begin{figure}
\begin{center}
\includegraphics[width=0.485\textwidth]{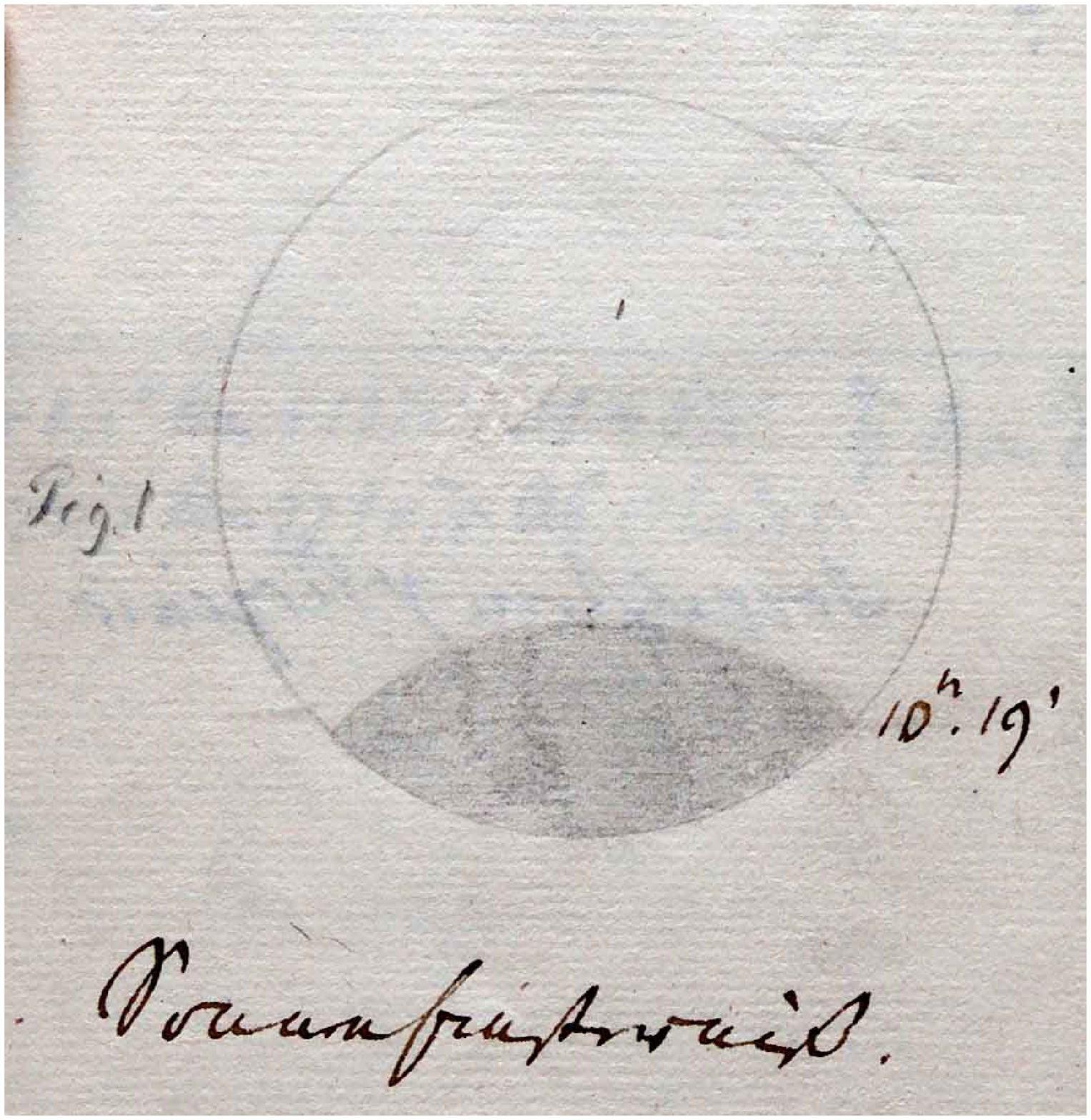}
\end{center}
\begin{center}
\includegraphics[width=0.485\textwidth]{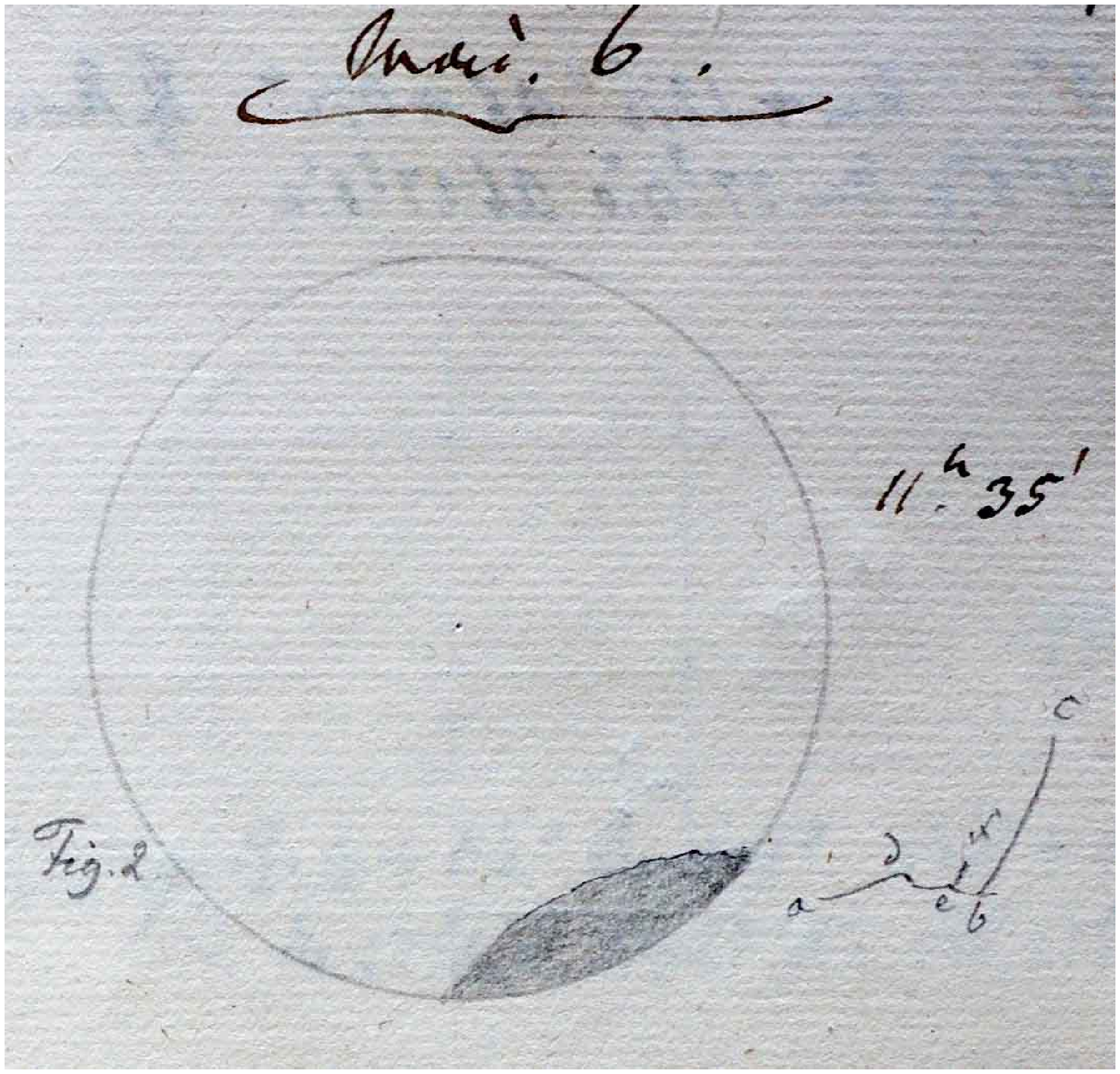}
\end{center}
\caption{Drawings of the solar eclipse of 1845 May~06 at 10\h 19\m\ 
and 11\h 35\m\ local time.}
\label{eclipse_1845}
\end{figure}

\begin{figure}
\begin{center}
\includegraphics[width=0.485\textwidth]{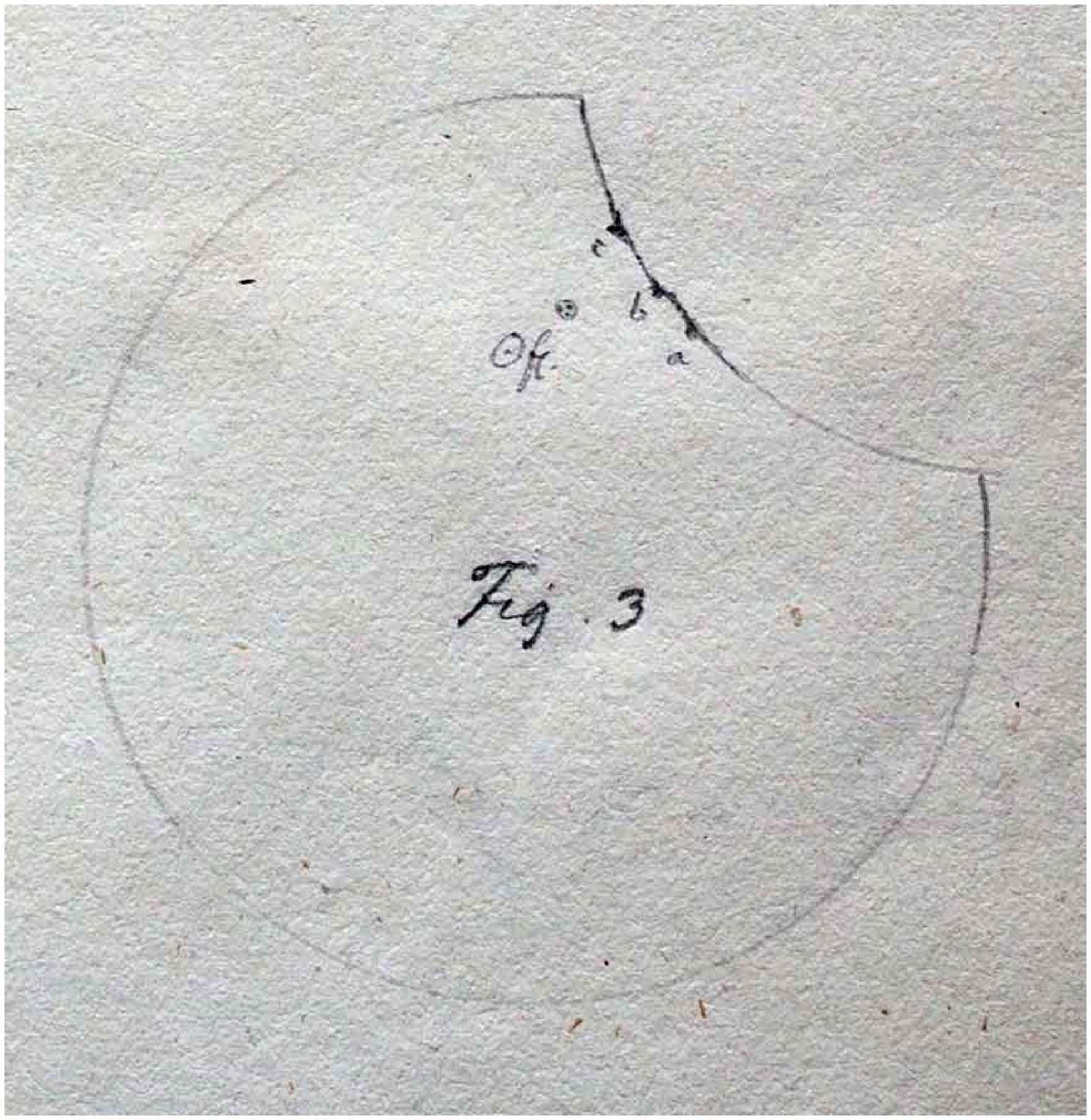}
\end{center}
\begin{center}
\includegraphics[width=0.485\textwidth]{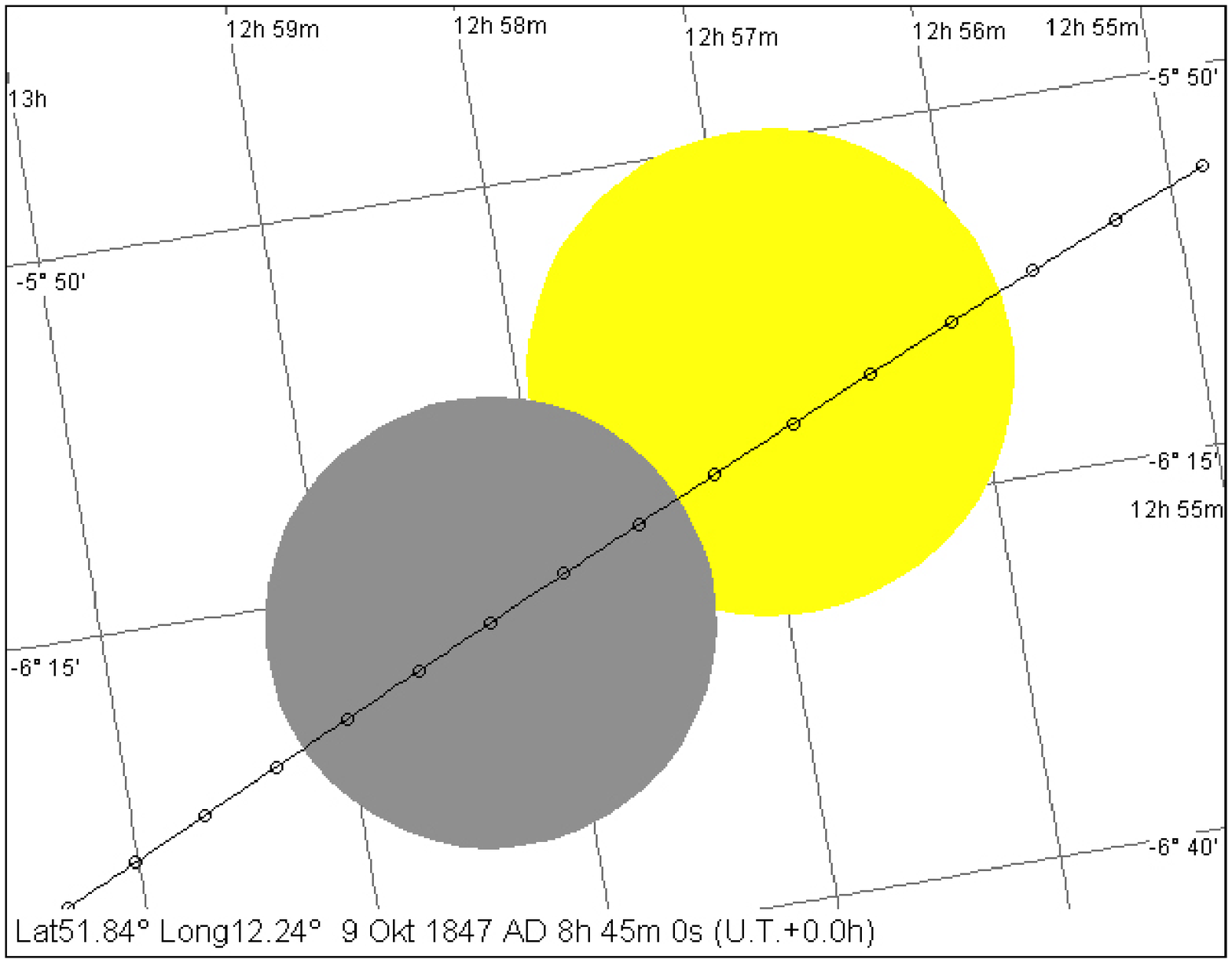}
\end{center}
\caption{Drawing of the solar eclipse of 1847 Oct~09. Bottom:
reconstruction of the lunar path with StarCalc by Alexander 
E. Zavalishin, http://homes.relex.ru/\~{}zalex/.}
\label{eclipse_1847}
\end{figure}

A final proof for the interpretation of the orientation as being
the view through a Keplerian telescope comes from the drawings of 
the solar eclipses. Figure~\ref{eclipse_1845} shows the eclipse
of 1845 May~06 with the Moon moving from left to right. The
path of the Moon was actually north of the solar disk center. The
eclipse of 1847 Oct~09 is shown in Fig.~\ref{eclipse_1847} with
a direct comparison with an ephemeris software.

It may perhaps be useful to use the observations of binary
stars to evaluate the quality of the telescopes. The following 
examples demonstrate the capabilities of the two main
telescopes; first the 3-1/2-foot Fraunhofer:
\begin{itemize}
\item The observation
 of $\sigma$~Ori shows  stars in a field of about $5'$ of which
 the two stars TYC 4771-1205-1 and TYC 4771-1204-1 are clearly
 separated. Their today's separation is $8.5''$.
\item Schwabe separated the two components of $\gamma$~Vir with his 
 3-1/2-foot telescope and a magnification of 168 times on 1827 
 Apr~29 which at the time had a distance of $2.1''$ and was 
 actually measured by a few observers (M\"adler 1838).
\end{itemize}
The 6-foot Fraunhofer telescope was of similar quality, and it
was actually used to draw a lunar atlas of highest resolution
by Lohrmann before. We just give two examples here:
\begin{itemize}
\item On 1829 Sep 04, Schwabe writes  ``Uranus disk-like with 6-foot, 
 144 times magnification.'' Uranus had a diameter of $3.7''$ at the
 time.
\item Schwabe separated the binary Castor with the 6-foot Fraunhofer
 on 1836 Jan~19 (magnification probably 216 times) and 1863 May~08
 (magnification 144 times); the distance was about $5.4''$ then.
\end{itemize}
It is strange though that Schwabe claims to have found a companion 
of Vega which does not exist. He writes on 1840 Aug~29 ``Vega in
Lyra fluttering strongly but the faint companion was visible'' and
on 1841 Nov~16 ``I see the faint companion very clearly''. However,
there is no star brighter than magnitude 12 in the $5'$ vicinity of
Vega.

The diffraction limits of the telescopes must have been between
$1''$ and $1.5''$ while the seeing in such a low-elevation city
place is at least as large. Visually, the seeing is not that 
harmful since very short clear moments can deliver information
on scales smaller than the seeing indicates. Small sunspots have
angular (celestial) diameters of about $2''$ and are resolvable
by Schwabe's telescopes. Since he occasionally describes the 
granulation, we can assume that the optical quality of his
telescopes was very high and resolving $2''$ structures was often
possible.

Schwabe also made a large number of measurements on sunspots.
In 1836, he started reporting on transit times and thus size measurements 
of the sun and sunspots with the threads in the 3-1/2-foot Fraunhofer and 
continued occasionally until November 1840. He obtained a transit
telescope made by Bobbe on 1840 Jun~20, but it was not until 
1840 Nov~05 when he actually measured the transit times of a
sunspot with this instrument. On 1840 Dec~21, he reports about
a ``Schraubenmicrometer'' (screw micrometer) which must be a 
device different from the micrometer with the transit telescope,
since Schwabe reported on both in parallel on 1840 Dec~28. He
gave calibrations of screw turns of the screw micrometer on
the first pages of his observing books of 1840, 1842, 1843, 1844, 
1845, 1846, 1847, and 1849. There are many measurements of transit times
of sunspots until 1842 which may be used to estimate the quality
of the actual drawings. The measurements ceased as the solar
activity minimum in 1843 approached.

\section{The times of day\label{times}}

\subsection{Time offset to UTC}
Which time reference was used? There was no common reference time
in Germany yet. Prussia started to use the Berlin time as a reference
in the middle of the 19th century, after telegraphic lines had been
installed. It is not clear whether the duchy of Anhalt-Dessau used the
same system. On 1830 Dec~23, the observation was
marked with ``12h mittl. Z.'' and later on, the abbreviation ``m.Z.'' 
appears until 1831 Feb~07, which translates into mean time. 
The abbreviation was omitted later, but
we may assume that Schwabe used the mean local time for his observations.
He obtained a sextant on 1830 Apr~29 with which he determined the time
of noon from ``corresponding solar elevations.'' It can be assumed that
the remaining deviations from the mean local time of his geographical
longitude are not more than a few minutes and will be negligible for
future analyses, given the limitations in plotting accuracy. Observations from
before 1830 may have larger timing errors though. Some observations
cannot be associated with any time of the day. These are also listed in
the inventory in Table~\ref{inventory} and are particularly frequent
until 1831.

We are now listing all the eclipses mentioned by Schwabe and try to deduce 
the reference time used in the observations he made, at least after 1830.
The eclipse times for Dessau were obtained from the NASA eclipse web site
by the Javascript Solar Eclipse Explorer
\footnote{http://eclipse.gsfc.nasa.gov/JSEX/JSEX-index.html}.
\begin{itemize}
\item The solar eclipse of 1826 Nov~29 does not help much, since no precise
times were reported for the two drawings given.
\item The eclipse of 1833 Jul~17 was not observed.

\item On the eclipse of 1836 May~15, Schwabe reports that his clock
still had some irregularities. He observed the first contact
at about 
14\h 56\m{} ``mean time''. The last contact was
seen at 
17\h 32\m 16\s. In reality, the first contact 
was at 
14\h 8\m 28\s~UTC while the last contact was at 
16\h 44\m 29\s~UTC 
corresponding to time differences of 48~min in both cases.

\item The eclipse of 1841 Jul~18 was reported to show its first contact 
at 
15\h 25\m 27\s\ including a time correction of 15 seconds. The drawing 
-- which is a few seconds after the first contact -- corresponds to 
the situation at 14h37m~UTC giving a time offset of 48~minutes.

\item There are no times given for the eclipse of 1842 Jul~18 because of clouds.

\item There are two drawings for the eclipse of 1845 May~06. The 
constellation of the lower panel of Fig.~\ref{eclipse_1845} was 
reached at $10^{\rm h}45.5^{\rm m} \pm 1$~min~UTC. Schwabe gave 
$11^{\rm h}35^{\rm m}$ for his drawing. The time offset is thus 
49.5~minutes. The drawing in the upper panel of Fig.~\ref{eclipse_1845} 
is not suitable because it shows the eclipse rather near the maximum 
phase ($9^{\rm h}55^{\rm m}$~UTC) when the distance between Sun and Moon 
changes very slowly.
%

\item For another eclipse on 1847 Oct~09, Schwabe determined the
last contact at 
9\h 51\m 31\s\ while the actual time was
9\h 3\m~UTC  indicating an offset of about 47.5~minutes.


\item There are also no exact times given for the eclipse of 1851 Jul~28.

\item The fourth contact of the eclipse of 1858 Mar~15 was observed
at 
15\h 17\m 20.3\s, the time obtained by the Solar Eclipse Explorer 
is 
14\h 27\m 24\s~UTC  corresponding to a time offset of a bit more 
than 49~minutes.

\item The observation of the eclipse of 1860 Jul~18 delivered a first
contact at 
14\h 43\m, whereas the Solar Eclipse Explorer obtains 
13\h 53\m 51\s~UTC, leading to a time offset of 49~minutes.

\item The eclipse of 1863 May~17 was not observed because of clouds.

\item There is a drawing of the eclipse of 1867 Mar~06, but no 
precise times were given.
\end{itemize}

The times are consistent with a mean local time for Dessau which
has an offset of 48.8~minutes to UTC, $t_{\rm UTC} = t_{\rm Schwabe}-48.8$~min. 
We recommend to use this offset for future analyses.

\subsection{Time referring to full-disk drawings}
In many cases, the observation consisted of several parts at 
different times of the day. While the precise knowledge of the
time is not essential for constructing the butterfly diagram,
it is important for measuring the differential rotation of the 
Sun. We need to associate the full-disk drawings with one of the
multiple observing times during the day. The verbal information
often does not say to which time the full-disk drawing refers.

Most of the observations starting with 1831 Sep~02 refer to 
12h of each day (after a one-week trip to Halle and Sandersleben, by the way). 
This is supported by many occasions when Schwabe either directly 
said the spots were plotted at noon, or verbal information which 
indicates that the full-disk drawing refers to 12h. A number of 
examples of these indications are given below.

\begin{itemize}
\item 1832 May~17:
``new spot 46 with penumbra not yet visible at 8:30 am.'' The full-disk
drawing contained this spot and indicates that it was made clearly
after 
8\h 30\m~am to when the rest of the description refers.

\item 1848 Aug~14:
``12h [\dots] Because of haze I could indicate only the general
positions of the spot groups.'' The drawing does indeed show only 
little spots as positions, no penumbrae or details, although a
detailed observation was reported for 6:30, when no drawing was 
obviously made.

\item 1853 May~26:
``7h pm [\dots] I saw a newly formed group.'' The new group was
not plotted in the full-disk drawing which is consistent with an
observation made at noon.


\item 1860 Jul~10:
``12h [\dots] 97 obtained five new subsidiary spots.'' These spots
are actually plotted in the drawing indicating it was made at noon
rather than at 5\h25\m~in the morning when he saw a single spot
with penumbra.

\item 1861 Mar~30:
While in the morning, Schwabe writes ``6 3/4 h [\dots] 42 a fine dot
and close to exiting'', he write at noon ``42 exited'' and does not
plot the spot in the full-disk drawing. This is a good indication
that he did not compile all the spots he saw during the course of
the day, but in fact drew what was visible at noon.

\item 1861 Dec~24:
``At noon I could see [\dots] the spots 201, 202 and the entering
spot 203.'' He writes at 2~pm that he saw ``199 not significantly
changed and 202 as two fine dots'' As these were not included in
the full-disk drawing, we may assume that Schwabe referred strictly
to noon for these drawings, at least during the last years of his
observations.

\item 1863 Jun~21:
``12 1/2 h [\dots] only 62 visible.'' Again, he did see another
group early in the morning but excluded it from the drawing made
at (or very near) noon.

\item 1864 Aug~13:
``7h am [...] 78 increased in size but close to exiting. 
12h [...] 78 exited definitively.'' This is yet another 
indication for the full-disk drawings referring to noon,
since group 78 was not plotted.

\item 1865 Jun~04:
``9 1/2h am [\dots] 51, 52, 54 not changed significantly.
12h [\dots] the spots could not be drawn. Weather cloudy during day.''
Indeed, the drawing is empty, most likely because of the weather.

\end{itemize}
Direct or indirect indications continue to suggest that drawings
were made preferably at noon until the end of 1867, if 12\h~is
mentioned among the various observing times Schwabe mentioned
during the course of each day. When the drawing was made at another 
time, it is usually mentioned clearly in the description.

We can also take the numerous disks without sunspots into account. 
These were typically made when no 12\h~observation is reported, so 
that means there was bad weather at noon, and no drawing could be 
made later. Schwabe drew the circle already in the morning but
clouds precluded the actual observation at noon.

\section{Coordinate system}
The majority of drawings show a cross of two diameter
lines or a net of six lines consisting of the two
diameter lines and a square with an edge length of about
64\% of the diameter. An example of such a grid is shown
in Fig.~\ref{grid1}. It is not likely to be meant as a 
golden ratio within the diameter, since it is clearly 
different from the 61.8\% and the construction would not
be any easier than other constructions. The lines must
have corresponded to hairs in the eye-piece, since Schwabe
writes on 1855 Jun~12: ``In the cross hairs of the 3-1/2-foot
[telescope], one hair was broken.'' (This is a footnote in
which he also noted the time when his wife Ernestine
Amalie `Malchen', n\'ee Moldenhauer, passed away.)

All tests with a superimposed heliographic grid delivered
a satisfactory spot distributions. The superposition includes
the inclination of the ecliptic against the celestial 
equator and the tilt of the solar rotation axis against the
ecliptic. We therefore assume, that whenever a net or two
axes are present in the drawing, we can be sure these were
aligned with the celestial equator and the direction to the 
celestial pole, as was the telescope itself. This is also
compatible with the description of Lohrmann's refractor
which clearly indicates a parallactic mount (Weichold 1985).

Some grids are rather skew, examples are 1840 Mar~23 and 
1840 Oct~31. The maximum deviation from $90^\circ$ of the 
vertical axis against the horizontal one is about $2.5^\circ$.
In case of a net, only the middle vertical is considerably
skew while the auxiliary lines near the limbs are better
aligned. There is no indication that the skewness was
drawn on purpose.

\begin{table*}
\caption{Solar drawings made by other people in assistance
to Schwabe.\label{other_persons}}
\footnotesize
\begin{tabular}{llrp{9.9cm}}
\hline
Year & Observer & Number & Dates of drawings \\
\hline
1831 & Alexander       & 0  & only seven verbal reports \\
1834 & Moldhauer       & 0  & only one verbal report \\
1837 & G\"otz          & 10 & 1837 06 07, 1837 06 08, 1837 06 09, 1837 06 12, 1837 06 13, 
                              1837 06 15, 1837 06 17, 1837 06 18, 1837 06 19, 1837 06 20\\
1837 & unknown 1       & 17 & 1837 09 15, 1837 09 16, 1837 09 20, 1837 09 21, 1837 09 22, 
                              1837 09 23, 1837 09 24, 1837 09 25, 1837 09 26, 1837 09 28, 
                              1837 10 01, 1837 10 04, 1837 10 05, 1837 10 08, 1837 10 10, 
                              1837 10 11, 1837 10 12\\
1837--1838 & unknown 2, possibly Krause 
                       &  8 & 1837 12 29, 1837 12 30, 1837 12 31, 1838 01 01, 1838 01 02,
                              1838 01 03, 1838 01 04, 1838 01 08\\
1838 & Krause          &  1 & 1838 01 18\\
1838 & Krause and Fritz&  1 & 1838 01 21\\
1838 & Krause, Fritz and Grube& 1 & 1838 01 22\\
1838 & Krause          &  8 & 1838 01 23, 1838 01 28, 1838 02 05, 1838 02 13, 1838 02 17, 
                              1838 02 18, 1838 02 20, 1838 02 22\\
1838--1839 & unknown 3 & 30 & 1838 05 06, 1838 08 12, 1838 08 14, 1838 08 15, 1838 09 02, 
                              1838 09 03, 1838 09 04, 1838 09 05, 1838 09 06, 1838 09 07,
                              1838 09 08, 1839 06 07, 1839 06 08, 1839 06 09, 1839 06 11,
                              1839 06 12, 1838 06 14, 1839 06 15, 1839 06 16, 1839 06 17,
                              1839 06 18, 1839 06 20, 1839 06 21, 1839 06 22, 1839 06 23,
                              1839 06 26, 1839 08 18, 1839 08 19, 1839 08 20, 1839 08 24\\
1846--1847 & unknown 4 & 23 & 1846 12 26, 1846 12 30, 1847 01 03, 1847 01 05, 1847 01 11,
                              1847 01 12, 1847 01 15, 1847 01 20, 1847 01 24, 1847 01 26,
                              1847 01 27, 1847 01 29, 1847 02 10, 1847 02 11, 1847 02 12,
                              1847 02 14, 1847 02 19, 1847 02 22, 1847 02 23, 1847 02 24,
                              1847 02 25, 1847 02 28, 1847 03 01\\
1860 & Marie           & 41 & 1860 01 05, 1860 01 06, 1860 01 08, 1860 01 09, 1860 01 09,
                              1860 01 10, 1860 01 12, 1860 01 13, 1860 01 14, 1860 01 19,
                              1860 01 20, 1860 01 21, 1860 01 25, 1860 01 27, 1860 01 28,
                              1860 01 30, 1860 01 31, 1860 02 01, 1860 02 02, 1860 02 03,
                              1860 02 09, 1860 02 11, 1860 02 14, 1860 02 20, 1860 02 21,
                              1860 02 28, 1860 02 29, 1860 03 01, 1860 03 02, 1860 03 08,
                              1860 03 10, 1860 03 11, 1860 03 12, 1860 03 13, 1860 03 14,
                              1860 03 15, 1860 03 17, 1860 03 18, 1860 03 19, 1860 03 21,
                              1860 03 27\\
1860 & Ferdinand       &  1 & 1860 01 07\\
1860 & Wilhelm Jahn    & 14 & 1860 02 26, 1860 04 02, 1860 04 03, 1860 04 04, 1860 04 05, 
                              1860 04 06, 1860 04 07, 1860 04 10, 1860 04 11, 1860 04 13, 
                              1860 04 15, 1860 04 15, 1860 04 16, 1860 07 28\\
1860 & Emilie W\"urzler&  5 & 1860 03 24, 1860 03 26, 1860 03 28, 1860 03 31, 1860 04 01\\
1863--1866 &Wilhelm Jahn&163& 1863 04 09, 1863 04 10, 1863 04 11, 1863 09 20, 1864 01 31,
                              1864 03 21, 1864 03 22, 1863 03 26, 1864 04 01, 1864 08 15,
                              1864 08 21, 1864 09 27, 1864 09 28, 1864 09 30, 1864 10 06, 
                              1864 10 07, 1864 10 09, 1864 10 10, 1864 10 14, 1864 10 15,
                              1864 01 17, 1864 10 18, 1864 10 19, 1864 10 22, 1864 10 23,
                              1864 10 24, 1864 10 25, 1864 10 26, 1864 10 27, 1864 10 31,
                              1864 11 01, 1864 11 02, 1864 11 03, 1864 11 06, 1864 11 07,
                              1864 11 08, 1864 11 09, 1864 11 10, 1864 11 11, 1864 11 13,
                              1864 11 14, 1864 11 15, 1864 11 19, 1864 11 20, 1864 11 23, 
                              1864 11 24, 1864 11 28, 1864 11 30, 1864 12 01, 1864 12 04,
                              1864 11 05, 1864 12 07, 1864 12 08, 1864 12 11, 1864 12 12,
                              1864 12 15, 1864 12 17, 1864 12 18, 1864 12 23, 1864 12 27,
                              1864 12 30, 1865 01 01, 1865 01 07, 1865 01 08, 1864 01 12,
                              1865 01 13, 1865 01 16, 1865 01 17, 1865 01 19, 1865 01 20,
                              1865 01 21, 1865 01 22, 1865 01 27, 1865 01 30, 1865 02 01,
                              1865 02 05, 1865 02 06, 1865 02 07, 1865 02 10, 1865 02 12,
                              1865 02 14, 1865 02 16, 1865 02 20, 1865 02 22, 1865 02 23,
                              1865 02 25, 1865 02 27, 1865 03 01, 1865 03 04, 1865 03 05,
                              1865 03 06, 1865 03 09, 1865 03 10, 1865 03 11, 1865 03 19,
                              1865 03 20, 1865 03 21, 1865 03 22, 1865 03 25, 1865 03 27,
                              1865 04 01, 1865 04 02, 1865 04 03, 1865 04 04, 1865 04 05,
                              1865 04 08, 1865 04 09, 1865 04 10, 1865 04 11, 1865 04 12,
                              1865 04 16, 1865 04 17, 1865 04 21, 1865 40 22, 1865 04 23,
                              1865 04 24, 1865 05 02, 1865 05 03, 1865 05 04, 1865 05 06,
                              1865 05 07, 1865 05 08, 1865 05 09, 1865 05 10, 1865 05 13,
                              1865 05 20, 1865 05 21, 1865 05 24, 1865 05 26, 1865 06 03, 
                              1865 06 08, 1865 06 09, 1865 06 16, 1865 06 18, 1865 06 20,
                              1865 07 01, 1865 07 04, 1865 07 08, 1865 07 23, 1865 08 07,
                              1865 08 19, 1865 08 23, 1865 08 26, 1865 09 05, 1865 09 26,
                              1865 09 27, 1865 09 28, 1865 10 01, 1865 10 02, 1865 10 08,
                              1865 10 16, 1865 10 29, 1866 04 16, 1866 04 17, 1866 07 25,
                              1866 07 26, 1866 09 02, 1866 09 19, 1866 09 23, 1866 10 04,
                              1866 10 06, 1866 12 28, 1866 12 31\\
1866 & Hartmann     &     7 & 1866 07 15, 1866 07 16, 1866 07 17, 1866 07 18, 1866 07 22,
                              1866 08 06, 1866 08 09\\
\hline
\end{tabular}
\end{table*}

A considerable number of drawings show no grid. These drawings
will pose problems to the measuring of the sunspots. We also
list the numbers of drawings without grid in Table~\ref{inventory}. 
Before 1831, the majority of drawings have no grid. Since the
observer has no reference lines except the solar limb, one
could think that these drawings are less accurate. Since the 
drift of the sunspots over the solar surface shows a very
consistent manifestation of the solar rotation, however, we may 
assume that the accuracy was yet fairly high. It may actually
be that a draft drawing with reference lines was first made, 
before the observation was copied into the actual logbook. But
there are no hints on how the observation was actually
conducted. An evaluation of the evolution of the accuracy
of the observing technique will be performed in the follow-up
paper on the positional measurements.

\section{Drawings by other persons}
Schwabe fell ill several times during his record of sunspot
observations, until 1831 because of fever attacks, and later
on due to podagra, a gout attack in the big toe. A number of
journeys also prevented Schwabe from observing occasionally.
A fracture of his leg prevented him from observing (Arendt 1925)
probably for several months in the 1860s to which the notes do
not directly refer. Several people then 
helped with observations. We do not know the full names of a 
few of them and in which relation they were to Schwabe.

A total of 329 drawings were made by other people which is 
3.9\% of the total number of drawings available. In all cases,
Schwabe added the group numbers afterwards, since the style
of writing them appears to be the same all the time. We can
therefore assume that Schwabe `approved' these drawings to
some extent. He comments ``Marie seems to have plotted the
spots incorrectly'' on 1860 Jan~14, although the drawing does
not look too bad, compared with the adjacent days. His 
critique underlines the precision he wished to achieve with
his observations.

All observations made by other persons are listen in Table~\ref{other_persons}.
We do not know the background of most of them. An exception
is Wilhelm Jahn who joined Schwabe's observations often over
the last years, first as a student and later as a teacher 
of mathematics and physics (Arendt 1925).

\begin{table}
\caption{Dates on which Schwabe claimed to have seen aurorae.
The one in 1852 is a report about a sighting from Berlin. The
uncertain ones are descriptions of atmospheric phenomena which
are fairly reminiscent of aurorae.
\label{aurorae}}
\begin{tabular}{ll}
\hline
Date                & Date    \\
\hline
1827 09 26          & 1847 12 17 \\
1831 01 07          & 1848 10 23\\
1831 08 10          & 1848 11 17\\
1833 09 17          & 1849 02 22\\
1834 11 03          & 1849 02 27\\
1837 02 18          & 1851 09 30\\
1837 11 12          & 1852 02 19 (not seen by Schwabe)\\
1838 09 15          & 1853 09 01 (uncertain)\\
1838 09 16          & 1859 08 28\\
1838 11 13          & 1859 09 03\\
1839 09 20 (deleted)& 1859 10 12\\
1840 12 21          & 1861 05 03 (uncertain)\\
1844 01 13          & 1861 12 02\\
1846 10 17          & 1862 12 14\\
1847 03 19          & 1863 10 08 (uncertain)\\
1847 10 13          & 1863 10 10 (uncertain)\\
1847 10 24          & \\
\hline
\end{tabular}
\end{table}

\section{Conclusions}
The solar observations by Samuel Heinrich Schwabe of 1825--1867 were
digitized. A total of 8486~full-disk drawings with sunspots are
available from 43~years. The drawings are precise enough to be
exploited for individual sunspot positions and areas. Times are
given as local times with an approximate offset of 49~min to UTC.
All drawings are upside-down as seen through a Keplerian telescope.
According to this preliminary evaluation, the accuracy of the
drawings will be particularly high for the periods of 1831--1867.

The measurements will be published in a follow-up paper. The
positions and spot sizes may not only be useful for the
butterfly diagram, but will also serve as a suitable record
for the analysis of group tilt angles, differential
rotation, possible hemispheric asynchrony, and possible active 
longitudes. Besides the drawings, there are verbal reports on
3699~additional days with which fairly precise lifetimes of 
spots can be obtained. The accuracy could also be enough for 
sunspot tracking and the derivation of meridional drifts as 
well as the evolution of the polarity separation.

Additionally, the possible sightings of aurorae can be compared
with other observational records of the time. The dates for which
Schwabe mentioned aurorae are listed in Table~\ref{aurorae}.

\acknowledgements
The author is very grateful to the Royal Astronomical Society,
London, for the permission to digitize the drawings and notes,
and in particular to Robert Massey and Peter Hingley for their 
support in this project. The author is also indebted to Anastasia
Abdolvand, Stela Frencheva, Jennifer Koch, and Christian Schmiel 
who helped utilizing the digital images in a usable format as 
well as with the inventory of the image set.


\end{document}